\documentclass[prb,preprint,byrevtex,floatfix,superscriptaddress]{revtex4}

\usepackage{amssymb,amsmath,graphicx,afterpage}
\usepackage{color}
\begin{document}
\title{\boldmath Chirality of Matter Shows Up via Spin Excitations\unboldmath}
\author{S. Bord\'acs}
\affiliation{Multiferroics Project, ERATO, Japan Science and Technology Agency (JST), Japan c/o The University of Tokyo, Tokyo 113-8656, Japan} \affiliation{Department of Physics, Budapest University of
Technology and Economics and Condensed Matter Research Group of the Hungarian Academy of Sciences, 1111 Budapest, Hungary}
\author{I. K\'ezsm\'arki}
\affiliation{Multiferroics Project, ERATO, Japan Science and Technology Agency (JST), Japan c/o The University of Tokyo, Tokyo 113-8656, Japan} \affiliation{Department of Physics, Budapest University of
Technology and Economics and Condensed Matter Research Group of the Hungarian Academy of Sciences, 1111 Budapest, Hungary}
\author{D. Szaller}
\affiliation{Department of Physics, Budapest University of
Technology and Economics and Condensed Matter Research Group of the Hungarian Academy of Sciences, 1111 Budapest, Hungary}
\author{L. Demk\'o}
\affiliation{Multiferroics Project, ERATO, Japan Science and Technology Agency (JST), Japan c/o The University of Tokyo, Tokyo 113-8656, Japan}
\author{N. Kida}
\affiliation{Multiferroics Project, ERATO, Japan Science and Technology Agency (JST), Japan c/o The University of Tokyo, Tokyo 113-8656, Japan}
\affiliation{Department of Advanced Materials
Science, The University of Tokyo, 5-1-5 Kashiwanoha, Chiba 277-8561, Japan}
\author{H. Murakawa}
\affiliation{Multiferroics Project, ERATO, Japan Science and Technology Agency (JST), Japan c/o The University of Tokyo, Tokyo 113-8656, Japan}
\author{Y. Onose}
\affiliation{Multiferroics Project, ERATO, Japan Science and Technology Agency (JST), Japan c/o The University of Tokyo, Tokyo 113-8656, Japan}
\affiliation{Department of Applied Physics, University of Tokyo,
Tokyo 113-8656, Japan}
\author{R. Shimano}
\affiliation{Multiferroics Project, ERATO, Japan Science and Technology Agency (JST), Japan c/o The University of Tokyo, Tokyo 113-8656, Japan}
\affiliation{Department of Physics, University of Tokyo,
Tokyo 113-0033, Japan}
\author{T. R\~o\~om}
\affiliation{National Institute of Chemical Physics and Biophysics, 12618 Tallinn, Estonia}
\author{U. Nagel}
\affiliation{National Institute of Chemical Physics and Biophysics, 12618 Tallinn, Estonia}
\author{S. Miyahara}
\affiliation{Multiferroics Project, ERATO, Japan Science and Technology Agency (JST), Japan c/o The University of Tokyo, Tokyo 113-8656, Japan}
\author{N. Furukawa}
\affiliation{Multiferroics Project, ERATO, Japan Science and Technology Agency (JST), Japan c/o The University of Tokyo, Tokyo 113-8656, Japan}
\affiliation{Department of Physics and Mathematics, Aoyama Gakuin University,
5-10-1 Fuchinobe, Chuo-ku, Sagamihara, Kanagawa 252-5258, Japan}
\author{Y. Tokura}
\affiliation{Multiferroics Project, ERATO, Japan Science and Technology Agency (JST), Japan c/o The University of Tokyo, Tokyo 113-8656, Japan}
\affiliation{Department of Applied Physics, University of Tokyo,
Tokyo 113-8656, Japan}
\affiliation{Cross-correlated materials group (CMRG) and correlation electron research group
(CERG), RIKEN Advanced Science Institute, Wako 351-0198, Japan}
\maketitle

\textbf{Right- and left-handed circularly polarized light interact differently with electronic charges in chiral materials. This asymmetry generates the natural circular dichroism and gyrotropy, also known as the optical activity.\cite{Barron2004,Berova2000} Here we demonstrate that optical activity is not a privilege of the electronic charge excitations but it can also emerge for the spin excitations in magnetic matter. The square-lattice antiferromagnet Ba$_2$CoGe$_2$O$_7$ offers an ideal arena to test this idea, since it can be transformed to a chiral form by application of external magnetic fields. As a direct proof of the field-induced chiral state, we observed large optical activity when the light is in resonance with spin excitations at sub-terahertz frequencies. In addition, we found that the magnetochiral effect,\cite{Baranova1979,Barron1984,Rikken1997} the absorption difference for the light beams propagating parallel and anti-parallel to the applied magnetic field, has an exceptionally large amplitude close to $100\%$. All these features are ascribed to the magnetoelectric\cite{Fiebig2005,Fiebig2005_2} nature of spin excitations as they interact both with the electric and magnetic components of light.}

Natural circular dichroism (NCD) and gyrotropy are observed respectively as the change in the ellipticity and the rotation in the polarization of the light transmitted through the medium. Since the sign of these quantities depend on the sense of handedness characteristic to the material, NCD is a common probe of chirality from submolecular to macromolecular level over a broad spectrum of the electromagnetic radiation as reviewed in Fig.~1 via the chiroptical study of proteins. Ultraviolet NCD spectroscopy of peptide-bond excitations is a well-established method for the determination of the secondary structure of proteins.\cite{Greenfield2007,Whitmore2008,Berova2000} Recent extensions of NCD spectroscopy to the infrared and X-ray photon-energy regions shed light on new signatures of chirality of matter via molecular vibrations \cite{Barron2004,Stephens2008} and core electron excitations.\cite{Alagna1998} Extending this context, the handedness of magnetic matter should also be detected in the NCD spectra of spin excitations, typically at gigahertz to terahertz frequencies, although seldom investigated.

When chirality is accompanied by magnetism, an intriguing optical cross effect, the magnetochiral dichroism \cite{Baranova1979,Barron1984} (MChD) emerges besides the conventional magnetically-induced circular dichroism (MCD). MChD is a directional dichroism and is measured as the absorption difference for unpolarized (or linearly polarized) light propagating parallel or antiparallel to the magnetization of the media. MChD has been found so far for visible light in metallic complexes,\cite{Rikken1997} molecular magnets,\cite{Train2008} inorganic crystals \cite{Rikken1998} and cholesteric liquid crystals \cite{Rikken2003} and is generally recognized as a weak effect. However, a recent study of the chiral phase of CuB$_2$O$_4$ proved that the magnetochiral dichroism can reach the order of unity ($\sim$100\%) in crystal field transitions of $d$ shell electrons.\cite{Arima2008}

Spin-waves (magnons) in conventional magnets are excited by the magnetic component of light and
they naturally show the MCD effect, while NCD and MChD effects are forbidden. A prerequisite to observe such phenomena in the long-wavelength region of light is the existence of dynamical coupling between spins and microscopic electric dipoles in the material.\cite{Barron2004} In multiferroic materials, where magnetization (M) and ferroelectric polarization (P) are interlocked, such cross-coupling is possible and was indeed observed as electrically excited magnetic resonance in several compounds including {\it R}MnO$_3$,\cite{PimenovREV2008,KidaREV2009} {\it R}Mn$_2$O$_5$, \cite{Sushkov2007} Ba$_2$Mg$_2$Fe$_{12}$O$_{22}$,\cite{Kida2009} BiFeO$_3$,\cite{Cazayous2008} and most recently in Ba$_2$CoGe$_2$O$_7$.\cite{Kezsmarki2011} However, the question of chirality has not been addressed so far in these systems. We predict that a magnetic field-induced chiral state with switchable handedness can emerge in magnets with non-centrosymmetric (but originally achiral) crystal structure as we demonstrate for Ba$_2$CoGe$_2$O$_7$. Moreover, the strong cross-resonances between electric- and magnetic-dipole active transitions existing at spin excitations in Ba$_2$CoGe$_2$O$_7$ can provide us with an excellent opportunity to observe gigantic NCD and MChD effects.

Ba$_2$CoGe$_2$O$_7$ has a tetragonal crystal structure where Co$^{2+}$ cations with S=3/2 spin form square-lattice layers stacked along the tetragonal [001] axis (see Fig.~2a). An antiferromagnetic order with tiny spin canting, as schematically shown in Fig.~2b, develops below T$_N$=6.7$\,$K.\cite{Zheludev2003} When the ferromagnetic component of the canted spins lie along the [110] axis, a ferroelectric polarization appears parallel to the tetragonal axis.\cite{Yi2008,Murakawa2010} The rotation of moderate magnetic fields (B$_{dc}\lesssim$1$\,$T) within the plane of the layers (tetragonal plane) rotates M accompanied by the sinusoidal modulation of P.\cite{Murakawa2010} While ferroelectricity vanishes in magnetic fields applied along the [100] or [010] axis, the corresponding spin patterns break all mirror-plane symmetries of the lattice, as visualized in Fig.~2b-c, and the crystal becomes chiral.\cite{Arima2008} We emphasize that although neither the spin nor the lattice structure alone is chiral, the whole spin-lattice system forms a chiral object. Due to this unique combination, the $\pi/2$ rotation of the magnetic field from [100] to [010] direction switches between the right- and left-handed forms of the magnetic crystal, since it is equivalent with a mirror plane reflection as demonstrated in Fig.~2b-c. This chiral state, which is the consequence of magnetoelectric coupling, is highly distinguished from the case of usual chiral molecules and crystal lattices with a given handedness where such a soft switching mechanism is hardly conceivable.

The nature of the spin excitations in the magnetically induced chiral state of Ba$_2$CoGe$_2$O$_7$ was
studied on high-quality single crystals with a typical thickness of 1\,mm by combining time-domain and Fourier transform terahertz spectroscopy. Besides the absorption coefficient, $\alpha\approx(-1/d)\ln(|\tilde{t}_{xx}|^2+|\tilde{t}_{xy}|^2)$, changes
in the polarization state of linearly polarized light due to NCD and gyrotropy were determined from the transmission components obtained by the time-domain method:
\begin{align}
\frac{\tan\theta+i\tan\eta}{1-i\tan\theta\tan\eta}=-\frac{\tilde{t}_{xy}}{\tilde{t}_{xx}},\nonumber
\end{align}
where $\tilde{t}_{xx}$ and $\tilde{t}_{xy}$ are the complex transmission coefficients
for the sample placed between parallel and crossed polarizers, while $\theta$
and $\eta$ are the polarization rotation and ellipticity, respectively.
In the terahertz frequency range we found two resonances located at f$\approx$0.5\,THz and
f$\approx$1\,THz. It was argued previously that the low-energy one resembles more a conventional
spin-wave excitation, while the unusual 1\,THz mode can be active with both the magnetic and electric components of the light and is viewed as magnetoelectric resonance.\cite{Kezsmarki2011,Miyahara2011}

We observed strong NCD and gyrotropy for each propagation direction and polarization configuration whenever
the static magnetic field, {\bf B$_{dc}$}, was pointing along [100] or [010] direction.
(The full chiroptical study is presented in the Methods section.)
When the light propagates along the [001] axis perpendicular to the static magnetic field and {\bf B$_\omega$}$\|${\bf B$_{dc}$}, as a typical case shown in Fig.~3a, both resonances can be excited. In high magnetic fields such as B$_{dc}$=7\,T, the low-energy magnon is clearly split into two
bands, while the magnetoelectric resonance at 1\,THz shows a less resolved splitting. The latter exhibits
giant gyrotropy and NCD characterized by $\theta\approx90^\circ$/mm, $\eta\approx45^\circ$/mm
around the resonance frequency. Such a large ellipticity indicates that the initially linearly polarized light becomes fully circularly polarized (disregarding additional linear dichroism effects) after passing through a 1\,mm thick crystal. The CD effect for the low-energy magnon mode is a factor of two smaller and has the opposite sign. These polarization phenomena are even functions of the magnetic field, thereby excluding any contributions from conventional MCD and providing the evidence of true chirality. The sign of the optical activity can be changed by
$\pi/2$ rotation of B$_{dc}$ from [100] to [010] as a direct proof for the
field-induced switching of the handedness of matter as shown in Fig.~3b. In fields applied along the intermediate [110] direction, by contrast, no optical activity appears, indicating the lack of chirality.

The existence of MChD was investigated in the chiral state
keeping {\bf B$_{dc}$} along [100] or [010] axis, while choosing the light
propagation to be parallel or antiparallel to B$_{dc}$. In this configuration, called Faraday geometry, the magnetically induced handedness is also manifested in strong NCD and gyrotropy (see Methods section).
To cover a broader spectral range with enhanced resolution, the experiments in Faraday configuration were performed by a Fourier transform terahertz spectrometer. The conventional magnetic resonance at $\sim0.5$\,THz is silent in this geometry and the magnetoelectric resonance at 1\,THz is split as discerned in Fig.~4a-b. In fields above 4\,T an additional strong mode enters from lower frequencies. This corresponds to the gapless Goldtstone mode  in the zero-field limit.\cite{Zheludev2003}

Depending on light propagation, parallel or antiparallel to the external field, the absorption shows a large difference,  $\Delta \alpha$$=$$\alpha^{+}$$-$$\alpha^{-}$, both in the region of the "Goldstone mode" and the magnetoelectric resonance as presented in Fig.~4c-d. Note that the "Goldstone mode" exhibits magnetochiral dichroism as large as $\Delta \alpha/\alpha_{max}\gtrsim70\%$ for the both polarizations in high fields, where $\alpha_{max}$$=$$max\{\alpha^+,\alpha^-\}$. Moreover, for the light polarization $B_{\omega}\parallel[010]$ shown in Fig.~4, the absorption nearly vanishes at the low-energy branch of the magnetoelectric resonance for the beam propagating antiparallel to the direction of $B_{dc}$. It means that the material becomes fully transparent at these frequencies when viewed from the direction of the magnetization, while a fairly large absorption is preserved for light traveling in the opposite direction, i.e. $\Delta \alpha/\alpha_{max}$ is close to $100\%$. As clear from Fig.~\ref{fig3}c-d, the optical magnetoelectric effect is appreciable even in low magnetic fields. Based on general symmetry arguments, we expect that MChD phenomenon persists even in lower B$_{dc}$ as long as the two-fold degeneracy of the antiferromagnetic ground state is lifted, which is a necessary condition for possible low-field applications.

NCD and MChD effect, emerging in Ba$_2$CoGe$_2$O$_7$ with large magnitude in the gigahertz-terahertz range, are observed for the first time for spin excitations. In the conventional view the spins only respond to the magnetic field of the light which induces an precessional magnetic moment described by the dynamical magnetic permeability $\hat{\mu}(\omega)$. However, in this multiferroic material, where the spins are coupled to microscopic electric dipoles, the relativistic spin-orbit interaction opens a channel to excite the spins via the electric field of the light,\cite{Miyahara2011} the so-called electromagnon process. Consequently, the dynamics of the spin system also induces oscillating electric polarization in the material expressed by the dielectric permittivity $\hat{\varepsilon}(\omega)$. Moreover, there is a strong interference of magnetic and electric responses when both the magnetic and the electric component of the light excite the same spin resonance. This cross-coupling, which is the optical analogue of the dc magnetoelectric effect and is represented by the dynamical magnetoelectric tensor $\hat{\chi}^{\rm me}$, is the origin of NCD and MChD effect at magnetic resonances. We have successfully described the magnetically induced chirality in the ground state of Ba$_2$CoGe$_2$O$_7$ and its spin excitation spectrum based on a microscopic spin model. Next, we solved the Maxwell equations using the dynamical response functions $\hat{\mu}(\omega)$, $\hat{\varepsilon}(\omega)$ and $\hat{\chi}^{\rm me}$ obtained from this theory. (These calculations are given in the Methods section.) The calculated absorption, gyrotropy/NCD spectra -- as approximated by the formula $\alpha^{\pm}=\frac{2\omega}{c} \mathfrak{Im}\{\sqrt{\varepsilon_{\gamma\gamma}\mu_{\delta\delta}}\pm\chi^{\rm me}_{\gamma\delta}\}$, $\theta+i\eta=\frac{i\omega}{2c}\{\chi^{\rm me}_{\gamma\gamma}+\chi^{\rm me}_{\delta\delta}\}$, respectively -- reproduce the experimental results in Fig.~3. Furthermore, the measured MChD spectra shown in Fig.~4c-d, which are the absorption difference for the light propagating parallel and and antiparallel to the magnetic field direction, shows an overall agreement with the theoretical curves $\Delta\alpha$$=$$\alpha^{+}$$-$$\alpha^{-}$$=\frac{4\omega}{c} \mathfrak{Im}\{\chi^{\rm me}_{\gamma\delta}\}$.

In conclusion, we have studied the spin dynamics in the easy-plane antiferromagnet Ba$_2$CoGe$_2$O$_7$ by terahertz spectroscopy. We demonstrated that via the dynamical or optical magnetoelectric effect present in multiferroic compounds the spin excitations can exhibit giant natural circular dichroism and gyrotropy -- what we call spin-mediated optical activity -- reflecting the magnetically induced and switchable handedness of this multiferroic material. As a consequence of chirality and broken time reversal symmetry, we have observed exceptionally large magnetochiral dichroism (nonreciprocal directional dichroism) of magnetoelectric spin excitations. Since simple spin orders, such as ferromagnetic or antiferromagnetic, can generate the optical magnetoelectric effect in compounds with non-centrosymmetric crystal structure, we expect that a broad variety of magnetic materials hosts such intriguing optical phenomena.

\textbf{Acknowledgements} We thank T. Arima and K. Penc for discussions. This work was supported by KAKENHI, MEXT of
Japan, by Funding Program for World-Leading Innovation R\&D on Science and Technology (FIRST) on ''Strong-
Correlation Quantum Science'', by Hungarian Research Funds OTKA PD75615, CNK80991, Bolyai
00256/08/11, T\'AMOP-4.2.1/B-09/1/KMR-2010-0002 and by the Estonian Ministry of Education and Research under Grant SF0690029s09, and Estonian Science Foundation under Grants ETF7011 and ETF8170.

\textbf{Author Contributions} S.B., I.K., T.R., U.N., D.Sz., L.D. performed the measurements and
analysed the data; H.M., Y.O. contributed to the sample preparation; N.K., R.S., T.R., U.N. developed the experimental set-up; S.M., N.F. developed the theory; S.B., I.K. wrote the manuscript; and Y.T., I.K. planned the project.

\textbf{Additional information} The authors declare no competing financial interests. Supplementary information accompanies this paper on www.nature.com/naturephysics.

\pagebreak

\begin{figure}[t!]
\includegraphics[width=4in]{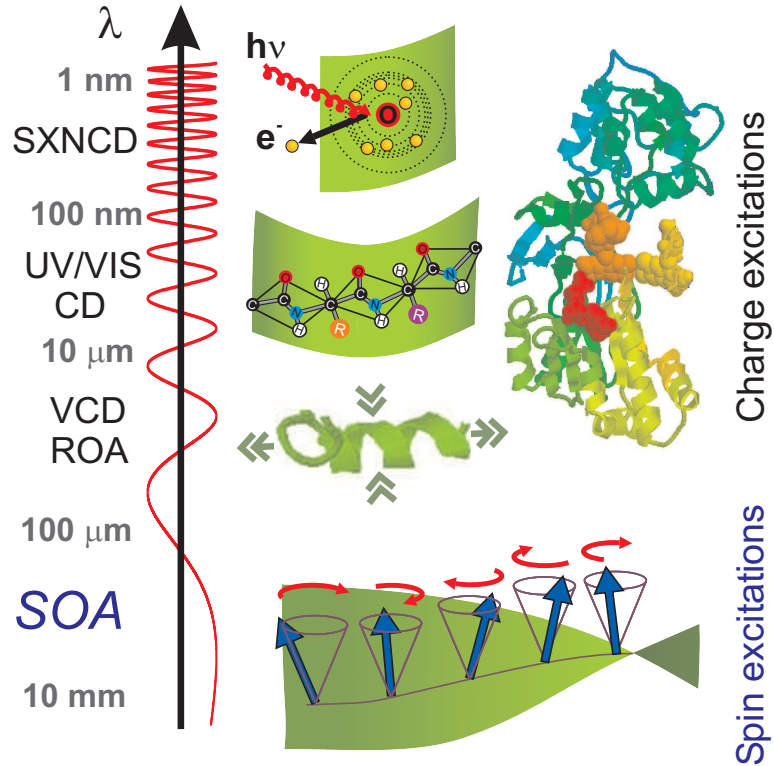}
\caption{\textbf{$\mid$ Chiroptical spectroscopy: An efficient probe of chirality both via electronic charge and spin excitations.} Depending on the wavelength, $\lambda$, the light interacts with various degrees of freedom and detects the handedness of matter at different levels. The soft X-ray natural circular dichroism (SXNCD) picks up chirality via the core electron excitations, while the circular dichroism in the UV and visible region (UV/VIS CD) probes it through transitions of valence electrons. Molecular vibrations are also sensitive to the handedness which is manifested in the vibrational circular dichroism (VCD) or the Raman optical activity (ROA). We predict that besides the charge excitations above, spin-wave excitations in the GHz-THz region ($\lambda$$\sim$$100\mu m$$-$$10mm$) of the electromagnetic spectrum can also probe the chirality of magnetic materials and show a spin-mediated optical activity (SOA).}
\label{fig0}
\end{figure}

\begin{figure}[h!]
\includegraphics[width=4.2in]{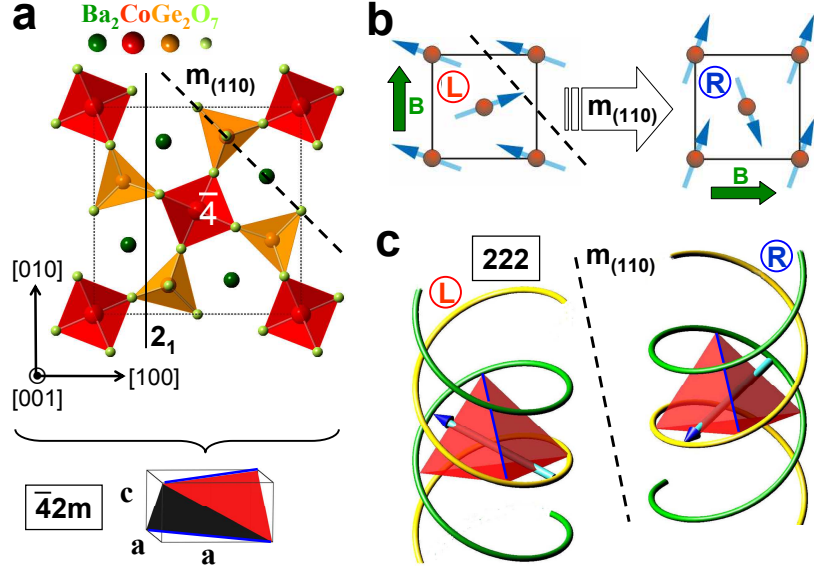}
\caption{\textbf{$\mid$ Main aspects of magnetically induced chirality.} \textbf{a,} The tetragonal crystal structure of  Ba$_2$CoGe$_2$O$_7$ with $\overline{4}$2m point group in the high-temperature paramagnetic phase. Fundamental aspects of the crystal symmetry are reflected by a single tetrahedron compressed in the [001] direction. (Its two shorter edges are indicated with blue lines.)
\textbf{b,} When the external magnetic field points along the [100] or [010] axis, the canted antiferromagnetic spin pattern breaks all mirror-plane symmetries of the lattice, hence, it makes the crystal chiral. Switching between the left-handed (L) and right-handed (R) enantiomers can be carried out by rotation of the magnetic field from [100] to [010] direction being equivalent to the m$_{(110)}$ mirror reflection.
\textbf{c,} A compressed tetrahedron together with a magnetic moment connecting midpoints of two opposite longer edges (representing the lattice symmetry and the spin system, respectively) form a chiral object having the same 222 symmetry as a double helix. In the mirror image, which has the opposite helicity, the tetrahedron remains unchanged while the magnetic moment being an axial vector is rotated by $\pi/2$.}
\label{fig1}
\end{figure}
\pagebreak

\begin{figure}[t!]
\includegraphics[width=4in]{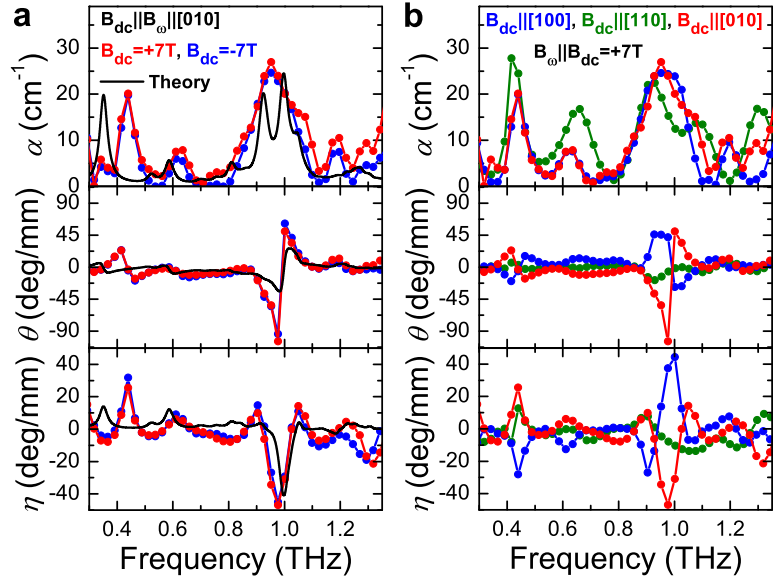}
\caption{\textbf{$\mid$ Absorption ($\alpha$), polarization rotation ($\theta$) and ellipticity ($\eta$) spectra
of the spin-wave modes} for light propagation along the [001] axis as measured by time-domain terahertz spectroscopy and calculated theoretically.
\textbf{a,} The modes observed at 0.5\,THz and 1\,THz in zero field show a clear (0.5\,THz$\rightarrow$0.43\,THz \& 0.62\,THz) and a less resolved splitting, respectively, in high magnetic fields parallel to the [010] axis. The large polarization rotation and ellipticity are even function of the magnetic field and have opposite sign for the two excitations.
\textbf{b,} When the external magnetic field direction is rotated from [100] to [010] axis,
both $\theta$ and $\eta$ change sign, while they are close to zero when the static magnetic field points
along [110] direction and the $m_{(110)}$ mirror-plane symmetry is restored. Thus, these polarization phenomena can be
clearly identified as spin-mediated optical activity in a chiral media.}
\label{fig2}
\end{figure}

\begin{figure}[h!]
\includegraphics[width=4.2in]{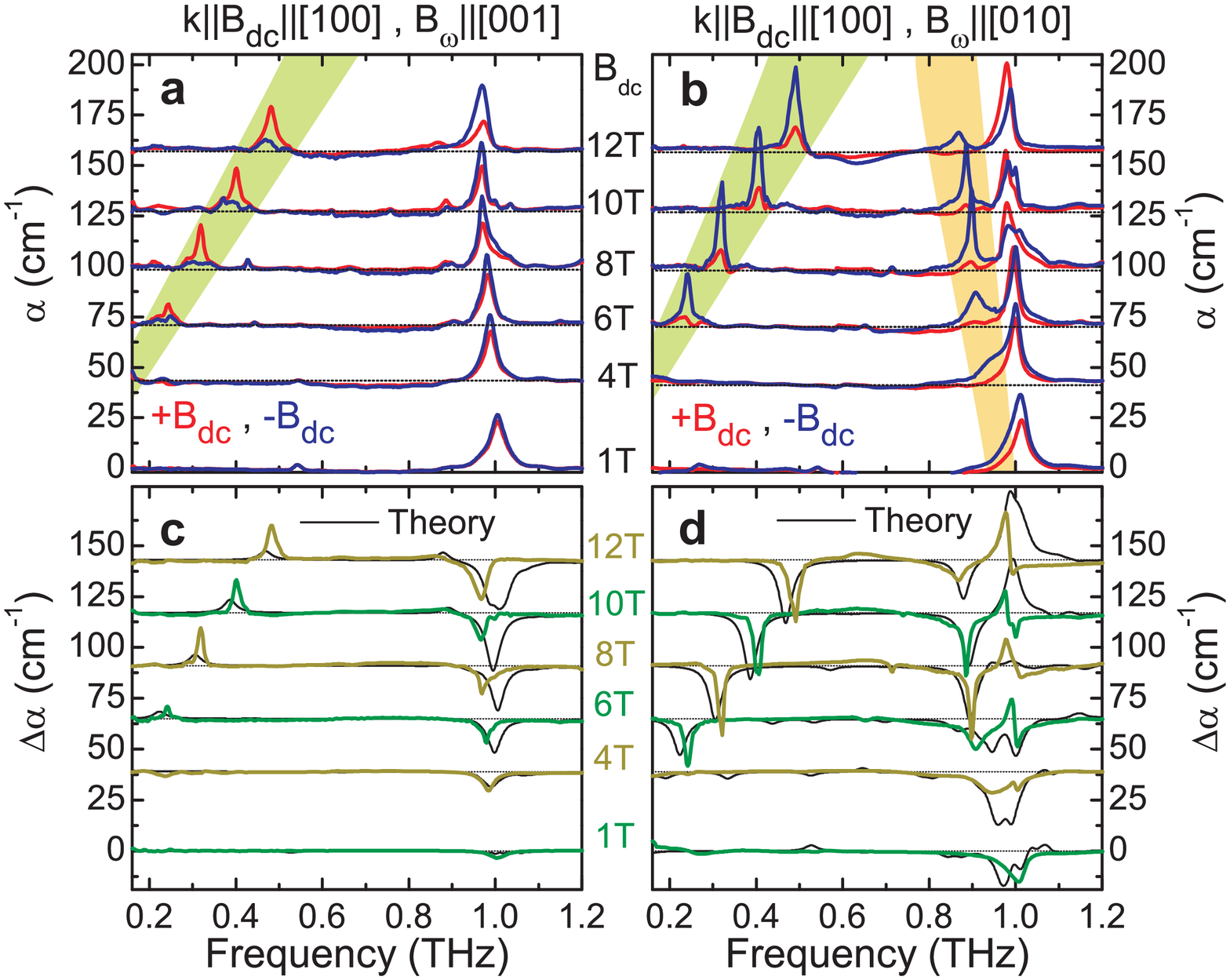}
\caption{\textbf{$\mid$ Absorption spectra in the field-induced chiral state.} Absorption is studied when the light beam propagates parallel and antiparallel to the external magnetic field $B_{dc}$$\parallel$[010] for the two possible polarization configurations. \textbf{a-b,} Spectra measured in different fields are shifted in proportion to the corresponding $B_{dc}$ values, while the red and blue curves correspond to the case of $+B_{dc}$ and $-B_{dc}$, respectively. Between the spin resonances the material is transparent as $\alpha$$\approx$$0$. The difference of the absorption coefficients for counter-propagating beams, the magneto-chiral dichroism, is huge both for the "Goldstone mode" and the magnetoelectric resonance at $\sim1$\,THz. $\Delta \alpha/\alpha_{max}\gtrsim70\%$ in the highlighted regions. \textbf{c-d,} The MChD effect, its sign and magnitude, is reproduced well for the both excitations by the theory.}
\label{fig3}
\vspace{0in}
\end{figure}


\begin{references}

\bibitem{Barron2004}Barron, L.D. {\it Molecular Light Scattering and Optical Activity}, (Cambridge University Press, Cambridge, 2004).
\bibitem{Berova2000}Berova, N., Nakanishi, K. \& Woody, R.W. (Eds.), {\it Circular Dichroism: Principles and Applications, second edn.}, (Wiley-VCH, New York, 2000.)
\bibitem{Baranova1979}Baranova, N.B. \& Zeldovich, B.Ya. Theory of a new linear magnetorefractive effect in liquids. {\it Molec. Phys.} {\bf 38}, 1085 (1979).
\bibitem{Barron1984}Barron, L.D. \& Vrbancich, Magneto-chiral birefringence and dichroism. {\it Molec. Phys.} {\bf 51}, 715 (1984).
\bibitem{Rikken1997}Rikken, G.L.J.A. \& Raupach, E. Observation of magneto-chiral dichroism. {\it Nature} {\bf 390}, 493 (1997).
\bibitem{Fiebig2005}Fiebig, M. Revival of the magnetoelectric effect. {\it J. Phys. D.: Appl. Phys.} {\bf 38}, R123 (2005).
\bibitem{Fiebig2005_2}Spaldin, N.A. \& Fiebig, M. The Renaissance of Magnetoelectric Multiferroics. {\it Science} {\bf 309}, 391 (2005).
\bibitem{Greenfield2007}Greenfield, N.J. Using circular dichroism spectra to estimate protein secondary structure. {\it Nat. Protoc.} {\bf 1}, 2876 (2007).
\bibitem{Whitmore2008}Whitmore, L. \& Wallace, B.A. Protein secondary structure analyses from circular dichroism spectroscopy: methods and reference databases. {\it Biopolymers} {\bf 89}, 392 (2008).
\bibitem{Stephens2008}Stephens, P.J., Devlin, F.J. \& Pan, J.-J. The determination of the absolute configurations of chiral molecules using vibrational circular dichroism (VCD) spectroscopy. {\it Chirality} {\bf 20}, 643 (2008).
\bibitem{Alagna1998}Alagna, L. {\it et al.} X-ray natural circular dichroism. {\it \prl} {\bf 80}, 4799 (1998).
\bibitem{Train2008}Train, C. {\it et al.} Strong magneto-chiral dichroism in enantiopure chiral ferromagnets. {\it Nature Mater.} {\bf 7}, 729 (2008).
\bibitem{Rikken1998}Rikken, G.L.J.A. \& Raupach, E. Pure and cascaded magnetochiral anisotropy in optical absorption. {\it \pre} {\bf 58}, 5081 (1998).
\bibitem{Rikken2003}Koerdt, C., Duchs, G. \& Rikken, G.L.J.A. Magnetochiral anisotropy in Bragg scattering. {\it \prl} {\bf 91}, 073902 (2003).
\bibitem{Arima2008}Saito, M., Ishikawa, K., Taniguchi, K. \& Arima, T. Magnetic Control of Crystal Chirality and the Existence of a Large Magneto-Optical Dichroism Effect in CuB$_2$O$_4$. {\it \prl} {\bf 101}, 117402 (2008).
\bibitem{PimenovREV2008}Pimenov, A., Shuvaev, A.M., Mukhin, A.A. \& Loidl, A. {\it et al.} Electromagnons in multiferroic manganites. {\it J. Phys.: Condens. Matter} {\bf 20}, 434209 (2008).
\bibitem{KidaREV2009}Kida, N. {\it et al.}, Terahertz time-domain spectroscopy of electromagnons in multiferroic perovskite manganites. {\it J. Opt. Soc. Am. B} {\bf 26}, A35 (2009).
\bibitem{Sushkov2007}Sushkov, A.B., Aguilar, R.V., Park, S., Cheong, S-W. \& Drew, H.D.
Electromagnons in Multiferroic YMn$_2$O$_5$ and TbMn$_2$O$_5$. {\it \prl} {\bf 98}, 027202 (2007).
\bibitem{Kida2009}Kida, N. {\it et al.} Electric-dipole-active magnetic resonance in the conical-spin magnet Ba$_2$Mg$_2$Fe$_12$O$_22$. {\it \prb} {\bf 80}, 220406 (2009).
\bibitem{Cazayous2008}Rovillain, P. {\it et al.} Electric-field control of spin waves at room temperature in multiferroic BiFeO$_3$. {\it Nature Materials} {\bf 9}, 975 (2010).
\bibitem{Kezsmarki2011}K\'ezsm\'arki, I. {\it et al.} Enhanced Directional Dichroism of Terahertz Light in Resonance with Magnetic Excitations of the Multiferroic Ba$_2$CoGe$_2$O$_7$ Oxide Compound. {\it \prl} {\bf 106}, 057403 (2011).
\bibitem{Zheludev2003}Zheludev, A. {\it et al.} Spin Waves and the Origin of Commensurate Magnetism in Ba$_2$CoGe$_2$O$_7$. {\it \prb} {\bf 68}, 024428 (2003).
\bibitem{Yi2008}Yi, H. T., Choi, Y. J., Lee, S. \& Cheong, S.-W. Multiferroicity in the square-lattice antiferromagnet of Ba$_2$CoGe$_2$O$_7$. {\it Appl. Phys. Lett.} {\bf 92}, 212904 (2008).
\bibitem{Murakawa2010}Murakawa, H., Onose, Y., Miyahara, S., Furukawa, N. \& Tokura, Y. Ferroelectricity Induced by Spin-Dependent Metal-Ligand Hybridization in Ba$_2$CoGe$_2$O$_7$. {\it \prl} {\bf 103}, 137202 (2010).
\bibitem{Miyahara2011}Miyahara, S. \& Furukawa, N. Theory of magnetoelectric resonance in two-dimensional $S=3/2$ antiferromagnet Ba$_2$CoGe$_2$O$_7$ via spin-dependent metal-ligand hybridization mechanism. {\it J. Phys. Soc. Jpn.} {\bf 80}, 073708 (2011).
\end{references}
\end{document}